\documentclass{article}

\usepackage{arxiv}

\usepackage[utf8]{inputenc} 
\usepackage[T1]{fontenc}    
\usepackage{hyperref}       
\usepackage{url}            
\usepackage{booktabs}       
\usepackage{amsfonts}       
\usepackage{amsmath}        
\usepackage{amssymb}
\usepackage{nicefrac}       
\usepackage{microtype}      
\usepackage{graphicx}
\usepackage[numbers,sort&compress]{natbib}
\usepackage{doi}

\title{Policy Description Language for Authorization\\ using Logic-Based Programming}

\author{
	\href{https://orcid.org/0000-0001-5596-282X}{\includegraphics[scale=0.06]{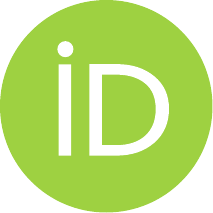}\hspace{1mm}Masaki Hashimoto} \\
	Graduate School of Information Security \\
	Institute of Information Security\\
	\And
	Mira Kim \\
	Graduate School of Information Security \\
	Institute of Information Security \\
	\And
	Hidenori Tsuji \\
	Graduate School of Information Security \\
	Institute of Information Security\\
	Advanced Institute of Information Technology\\
	\And
	Hidehiko Tanaka \\
	Graduate School of Information Security \\
	Institute of Information Security \\
}

\renewcommand{\shorttitle}{Policy Description Language for Authorization using Logic-Based Programming}

\hypersetup{
pdftitle={Policy Description Language for Authorization using Logic-Based Programming},
pdfauthor={Masaki Hashimoto, Mira Kim, Hidenori Tsuji, Hidehiko Tanaka},
pdfkeywords={access control, policy description language, logic programming, Datalog, SELinux, defense-in-depth},
}

\begin{document}
\maketitle

\let\thefootnote\relax\footnotetext{
	This document is an English translation of the authors' paper originally published in Japanese: M.~Hashimoto, M.~Kim, H.~Tsuji, and H.~Tanaka, ``Policy Description Language for Authorization using Logic-Based Programming,'' IPSJ Journal, Vol.51, No.9, pp.1682-1691, Sep. 2010.
	\copyright~Information Processing Society of Japan (IPSJ).
}

\begin{abstract}
Recently, with the impossibility of eradicating the vulnerabilities of information systems, we must prepare for the occurrence of the security incident by the multi-layer defense called the Defense-in-Depth strategy. In the multi-layer defense, it is important to authorize accesses in fine-grained granularity to compose each layer effectively, and many access control models are proposed to follow them. However, policy description languages proposed so far cannot express the models appropriately in proper granularity. In this paper, we propose a policy description language which can designate many kinds of conditions for access control, such as the dynamic status of an application process, as an element of decision data, and implement it in Datalog. Using the proposed language, we compose the policy of SELinux, which is a major implementation achieving the multi-layer defense, and we confirm the advantages of the proposed language by evaluating its validity and expressiveness.
\end{abstract}

\keywords{Access control \and Policy description language \and Logic programming \and Datalog \and SELinux \and Defense-in-Depth}

\section{Introduction}
\label{sec:Introduction}
Because recent information systems are large-scale and complex, it is difficult to build a system that is free from vulnerabilities. Therefore, as a realistic security measure, it is effective to design systems on the premise that vulnerabilities exist and to adopt the Defense-in-Depth strategy~\citep{DID}. Multi-layer defense partitions an information system strictly so that, even if a zero-day attack by malware or a virus is executed and intrusion into the system is permitted, attacks against important information assets can be delayed and the scope of damage can be localized.

The effectiveness of multi-layer defense increases as the system interior is partitioned more finely by fine-grained mandatory access control; however, this requires describing a huge number of access control rules as a policy. On the other hand, if the granularity of the partitions is made coarse to reduce the amount of access control rules, the direct scope of damage when a security incident occurs expands, and the effect of delaying damage propagation also weakens, thereby offsetting the effectiveness of multi-layer defense. Hence, to apply multi-layer defense effectively, a device for describing a huge policy concisely is required. However, in existing implementations of multi-layer defense such as SELinux~\citep{Loscocco:715771}, the main approach is to describe each access control rule directly and individually, so the amount of description tends to become huge as the granularity becomes finer, and it is difficult to grasp the behavior of the policy as a whole without using a dedicated verification tool.

This research proposes a policy description language that, by expressing access control rules as a logic program, enables inheritance of attributes and descriptions structured as subroutines, and aims to solve this problem. The language is designed in its essentials by defining the syntax, the formal semantics, and the inference rules of declarative statements for describing each access control rule and of authorization queries for querying the policy composed as a set of declarative statements; it is implemented using Datalog~\citep{Ceri:1989p4021}, a subset of Prolog~\citep{Warren:806939}. Furthermore, in this paper, after presenting a method for structurally describing concrete access control models using the language, we construct an experimental system that actually describes the policy of SELinux, and we evaluate the validity and the expressiveness of authorization decisions.

The remainder of this paper is organized as follows. First, Section~\ref{sec:Background} explains multi-layer defense as a strategy that presupposes the existence of vulnerabilities, introduces SELinux as an implementation example of multi-layer defense and recent research trends related to policy description as related work, and then states the purpose of this research. Next, Section~\ref{sec:Overview} presents the grammar, semantics, and inference rules as the essential design of the language, and explains the implementation method using Datalog. Section~\ref{sec:Expression} then explains concrete methods for describing access control models using the language, and Section~\ref{sec:Evaluation} describes the evaluation experiments of the language and their discussion. Finally, Section~\ref{sec:Conclusion} concludes the paper.

\section{Background and Purpose of This Research}
\label{sec:Background}
This chapter first explains, as the background of the research, multi-layer defense as a protection method for information systems, focusing on its relationship with vulnerabilities. It also takes up SELinux as an implementation example, and explains the method of realizing multi-layer defense using fine-grained mandatory access control and the policy description that becomes an issue in doing so. After that, it presents recent research related to policy description and finally explains the purpose of this research.

\subsection{Vulnerabilities and Multi-Layer Defense}
Because recent information systems are large-scale and complex, it is difficult to build a system free from vulnerabilities. Although mechanisms to prevent the introduction of vulnerabilities have been refined through advances in software analysis techniques and development methods, attacks on information systems often actively search out the weakest part and intentionally trace and execute that path; therefore, even a minute vulnerability that could normally be ignored may put the system into a fatal state. For this reason, from a security standpoint, as long as vulnerabilities cannot be eradicated, it is necessary to take into account the possibility that an attack succeeds, even if that possibility is low. Accordingly, as a realistic security measure that takes this into account, it is effective to design the system on the premise that vulnerabilities exist and to adopt the Defense-in-Depth strategy. The multi-layer defense strategy partitions an information system strictly so that, even if a zero-day attack by malware or a virus is executed and intrusion into the system is permitted, attacks against important information assets can be delayed and the scope of damage can be localized.

In the future, it is expected that information systems will become larger and more complex than before through cloud computing, Web services, and Grid computing; as a result, eradicating vulnerabilities from systems will become even more difficult. Therefore, to build a truly trustworthy computing environment, building a system on the premise that vulnerabilities exist is realistic, and for that purpose the realization of a multi-layer defense strategy is desired.

To realize multi-layer defense, one first determines an access control model that defines the policy for composing the layers. Various access control models have been proposed, such as RBAC~\citep{Ferraiolo:1992p5192}, DTE~\citep{Tidswell:1997p5200}, and HBAC~\citep{Banerjee:2005p5185}, and each has its advantages and disadvantages, so an access control model must be chosen according to the characteristics of the system to be realized. Once the access control model is determined, the next thing needed is to concretely describe the access control rules according to that policy, which is realized by a policy description language. A policy description language must be able to express the access control model without excess or deficiency, and it is desirable for it to be intuitively understandable to the person describing it. The last thing needed is a mandatory access control mechanism, also called a reference monitor. To enforce, without omission and reliably, the access control rules described by the policy description language, the mandatory access control mechanism captures and mediates all accesses. In recent years, secure operating systems have been implemented that use the OS as an unavoidable path when a process accesses a resource, thereby realizing this.

\subsection{Related Work}
\label{sec:works}
This section first explains SELinux as an implementation that realizes multi-layer defense, and uses it as an example to present the issues related to policy description. Next, it explains recent research trends related to policy description and states their relation to this research.

\subsubsection{SELinux's Method of Realizing Multi-Layer Defense and Its Issues}
\label{sec:selinux}
SELinux is a representative secure OS whose development is being actively advanced by an open-source community centered on the U.S. National Security Agency. SELinux is an implementation of the Flask security architecture~\citep{Spencer99theflask} for Linux, and it consists of a security server, an object manager, and an access vector cache. In its Linux implementation, it seizes the in-kernel control related to access control by hooking system calls, and SELinux enforces access control according to its own policy.

Because SELinux can, in principle, make access control decisions for all system calls, it is possible to realize the RBAC model and the DTE model at the granularity of the system call level. On the other hand, however, to perform fine-grained access control, the policy that serves as the control rules must naturally also be described in a fine-grained manner, which is an extremely difficult task for humans.

For example, in SELinux a policy is composed by listing access control rules such as the following.

\begin{quote}
allow $acct\_t$ $unconfined\_t$:$fd$ $use$;\\
allow $acct\_t$ $init\_t$:$process$ $sigchld$;\\
allow $acct\_t$ $init\_t$:$process$ $signull$;\\
allow $acct\_t$ $rpm\_t$:$fd$ $use$;
\end{quote}

In this example, each line expresses one access control rule. For instance, in the first line, the first item \textit{allow} indicates that the line is a rule that grants permission. The second item is the information of the subject to which this rule applies, and the third and subsequent items indicate the information of the object and the operation. In SELinux, even the simplified policy installed by default for individual users lists nearly 40,000 lines of access control rules in the grammar above. Therefore, even if each access control rule can be understood line by line, it is difficult to judge what actually happens with the policy composed as their totality. Moreover, in recent years, implementations have been proposed that expand the scope of SELinux's mandatory access control rules from resources of the OS layer to elements of the application layer and elements of other systems~\citep{SESQL,PMS}, so it can be inferred that appropriate policy description becomes increasingly difficult.

\subsubsection{Recent Research Situation Concerning Policy Description Languages}
\label{sec:researches}
SDSI~\citep{Abadi:353678} and SPKI~\citep{RFC2693} are frameworks for evaluating the trustworthiness between systems by exchanging certificates when performing access control. SDSI constructs its logic using a predicate called ``speaks-for,'' and in SPKI predicates can be encoded as tags within a declaration. Both SDSI and SPKI have the feature that policies can be expressed by a syntax similar to English; however, because SDSI uses only ``speaks-for'' as a predicate, it has the limitation that predicates suited to an application cannot be defined. SPKI can use multiple predicates, but because it is constrained to predefined tags and cannot handle variables, it likewise has the issue that it is difficult to define appropriate predicates. Similarly, PolicyMaker~\citep{Blaze:728332} and KeyNote~\citep{Blaze:720552} are frameworks that have the feature of being able to define various states required for authorization decisions, but because there are strict restrictions on predicate definition, predicates suited to an application cannot be used. Since these languages focus on trust management between domains, they are effective even with a somewhat restricted set of predicates, but when used as authorization decision languages it is difficult for them to handle the various access control models that are actually used. Therefore, in this research we examined a method that makes it possible to describe predicates with arbitrary meanings by defining a syntax that designates user-defined operation contents as part of the predicate.

SecPAL~\citep{Becker:1270639} is a policy description language based on a constraint logic language, in which an authorization request is granted when a query against a set of clauses succeeds. SecPAL's grammar is close to natural language, its semantics are composed of three inference rules, and by supporting negative queries, recursive predicates, delegation of authority with a specifiable number of times, and various other restrictions, it can express many access control models in a general-purpose manner. Lithium~\citep{Lithium}, like SecPAL, is a language that describes policies using inference based on first-order predicate logic, and it realizes inference including negation by restricting recursive expressions. These languages share with this research the point of expressing policies through a logic-type language and inference; however, the former in particular is research that presupposes, in relatively large-scale systems, composing policies by modularizing them per administrative domain, and that focuses on a description method for flexibly delegating authority among distributedly managed domains, while the latter concentrates on formally and correctly inferring logical negation. This research differs from these in that it aims to improve the efficiency of the description method itself by raising the level of abstraction of the description of authorization decision rules.

\subsection{Purpose of This Research}
\label{sec:purpose}
The purpose of this research is to propose a policy description language that, by describing access control rules as a logic program, supports the inheritance of attributes and the subroutinization of authorization procedures, and thereby to solve the problem of the complexity of policy description through a flexible structured description method.

Conventional policy description methods, such as SELinux exemplified in Section~\ref{sec:works} and XACML~\citep{XACML}, describe each access control rule independently and directly designate each of their elements, so they had problems of readability and maintainability. This problem becomes larger as the system interior is partitioned more finely in order to improve the effectiveness of multi-layer defense, because the granularity of the required access control rules also becomes finer. On the other hand, making the granularity of access control rules coarse can reduce the amount of description, but in that case the direct scope of damage when a security incident occurs expands, and the effect of delaying damage propagation also weakens, thereby offsetting the effectiveness of multi-layer defense. In addition, when, with the future spread of the cloud and Grid, cooperative processing among multiple systems becomes common, and the whole is controlled by a policy rather than each system arbitrarily defining its own access control rules, it is easy to imagine that the amount of description will explode.

Therefore, to counter these problems, various studies on policy description such as those shown in Section~\ref{sec:researches} have been conducted in recent years; however, existing research has insufficiently examined description methods that reduce the amount of policy description itself. Hence, this research focuses particularly on this point and examines a method for describing efficiently by raising the level of abstraction of policy description. At the same time, by an experiment in which a policy that is actually used is described with the proposed method, we quantitatively evaluate and confirm that the policy described as a program operates logically correctly to make authorization decisions and that the amount of description is reduced.

By describing a policy structurally as a logic program, we can expect to naturally apply mathematical-logical underpinnings to the policy description. With this, for example, we can expect to incorporate into the description method itself a policy verification function that is normally realized as a separate, independent application program, and by using the verification result as a condition of access control, to eliminate conflicts of access privileges in advance; and to grant a minimal set of privileges according to the content of work to a proxy process that processes work requested from another system. In addition, we can consider a function that controls how new rules are added or existing rules are changed by auxiliary rules, and a function in which the authorization decision mechanism changes the arguments used for the authorization decision according to the situation. Furthermore, by verifying whether the access control target holds a particular attribute, we can expect to perform access control based on dynamically transitionable information such as ``being stopped'' or ``being tainted,'' thereby preventing unintended information flows in advance.

\section{Design and Implementation of the Policy Description Language}
\label{sec:Overview}
The language is composed of policy declarative statements for setting access control rules, and authorization queries for querying the policy about whether access is permitted. This chapter first formally defines the syntax and meaning of each of these and describes the inference rules applied to them, thereby explaining the design of the language. After that, it explains the implementation of the language using Datalog, a subset of Prolog.

\subsection{Design of the Policy Description Language}
\label{sec:design}
This section, as the design of the policy description language, first defines the syntax of the policy declarative statements and the authorization queries, and then explains their formal meaning and inference rules. The language is devised to be easy to understand through a syntax similar to English, its semantics, and simple inference rules.

\subsubsection{Syntax of Policy Declarative Statements and Authorization Queries}
The language composes a policy as a set of declarative statements (DS:~\textit{Declarative Statements}) described by the following syntax, and during access control it queries the policy about whether access is permitted by means of an authorization query (Q:~\textit{Authorization Query}).

\begin{quote}
\begin{tabbing}
$fact$ \= ::= \= \kill
DS \> ::= \> \textbf{Policy specifies} $fact$ \textbf{if} $fact_1,\ldots,fact_n$ $(n\ge 0)$\\
Q \> ::= \> \textbf{Policy specifies} $fact$ $\mid$ $\exists x$(Q)\\
S \> ::= \> $E$ \textbf{tagged} $A$\\
V \> ::= \> \textbf{is permitted to} $O$ S $\mid$ \textbf{is forbidden to} $O$ S $\mid$ \textbf{moves to} S \textbf{with} $O$\\
I \> ::= \> \textbf{inherits} $A$\\
$fact$ \> ::= \> S V $\mid$ $A$ I $\mid$ S
\end{tabbing}
\end{quote}

The language is based on predicate logic. Among the above, those shown in bold typeface represent predicates, which are the basic elements of the syntax. $E$, $O$, and $A$ are terms denoting, respectively, the access subject and its target, the operation content, and the attribute. It is assumed that one domain to which the term belongs is associated with every term. Those shown in italic lowercase denote variable terms, and substitution of a term for a variable term is permitted only when the domain of the variable term and that of the term agree. Also, $fact_i$ is of the same kind as $fact$, with a suffix attached to distinguish each one. S is an access subject or target with an attribute attached; in the language, an access control rule is expressed by giving the predicate V to a subject part S. I is another element that can serve as a predicate, in which case the subject part is only $A$. The intuitive meaning of each predicate is as follows: by ``\textbf{is permitted to}'' and ``\textbf{is forbidden to},'' the subject part $\textrm{S}_{1}$ permits or forbids the operation $O$ to a certain target $\textrm{S}_{2}$; and by ``\textbf{moves to},'' the subject part $\textrm{S}_{1}$ transitions to $\textrm{S}_{2}$ triggered by $O$.

\subsubsection{Formal Meaning of the Syntax and Inference Rules}
\label{rules}
In the language, when an authorization query Q is given to a policy DS, which is a set of declarative statements, the interpretation of the authorization query is determined by the relation DS,$\theta\vdash$ Q shown below. The truth or falsity of Q is interpreted respectively as permission or denial at the time of the authorization decision, but when Q contains a variable term, all responses that make Q true are returned. Here $\theta$ is a mapping that associates a variable term with another variable term or a constant term, and $fact\theta$ is the result of applying $\theta$ to $fact$. Also, $\theta_{-x}$ denotes a mapping whose domain is dom($\theta$)$-\{x\}$, and vars(X) denotes the variable terms present in clause X.

\begin{quote}
DS,$\theta\vdash$ \textbf{Policy specifies} $fact$ iff DS $\models$ \textbf{Policy specifies}
$fact\theta$,\\
\hspace*{3em} and dom($\theta$) $\supseteq$ vars(\textbf{Policy specifies} $fact$)\\
DS,$\theta_{-x}\vdash\exists x$(Q) iff DS, $\theta\vdash$ Q
\end{quote}

Next, the inference rules applied to declarative statements and authorization queries are defined as follows. An inference rule derives another, new syntactic construct by being automatically applied to a particular syntactic construct, and it can reduce the amount of description of declarative statements. Inference rule (a) automatically derives a $fact\theta$ without variable terms when there is a declarative statement with a conditional clause and a $\theta$ that makes vars($fact\theta$)=0 for each $fact$ of that conditional clause. Note that vars($fact\theta$)=0 indicates that no variable term exists in $fact\theta$. Inference rule (b) automatically derives, when there is a construct in which attribute $A_{2}$ inherits attribute $A_{1}$ and a construct that declares $\textrm{V}_{1}$ for a variable $x$ tagged with attribute $A_{1}$, a construct that declares $\textrm{V}_{1}$ also for the variable $x$ tagged with attribute $A_{2}$.

\paragraph{Inference rule (a).}
\begin{equation*}
\frac{
\begin{array}{c}
(\textbf{Policy specifies}\ fact\ \textbf{if}\ fact_1,\ldots,fact_k)\in \textrm{DS}\\[2pt]
\textrm{DS}\models\textbf{Policy specifies}\ fact_i\theta\ \textrm{for all}\ i\in\{1..k\}\\[2pt]
\textrm{vars(}fact\theta\textrm{)}=0
\end{array}
}{
\textrm{DS} \models \textbf{Policy specifies}\ fact\theta
}
\end{equation*}

\paragraph{Inference rule (b).}
\begin{equation*}
\frac{
\begin{array}{c}
\textrm{DS}\models\textbf{Policy specifies}\ A_{2}\ \textbf{inherits}\ A_{1}\\[2pt]
\textrm{DS} \models \textbf{Policy specifies}\ x\ \textbf{tagged}\ A_{1}\ \textrm{V}_{1}
\end{array}
}{
\textrm{DS} \models \textbf{Policy specifies}\ x\ \textbf{tagged}\ A_{2}\ \textrm{V}_{1}
}
\end{equation*}

\subsection{Implementation of the Policy Description Language}
\label{sec:Details}
This section explains the implementation of the language based on Datalog. Datalog is a logic programming language that is a subset of Prolog, and the language is realized by translating declarative statements and authorization queries into a Datalog program. By expressing access control rules using logic programming, the language can provide a mathematical-logical underpinning to policy description while enabling declarative description.

\subsubsection{Overview of Datalog and Its Extended Implementation XSB}
Datalog is a language based on logic programming, and it is in particular a rule-based language designed to exchange information with large-scale databases. That is, it provides an interface for accessing data directly and supports the exchange of information by a rule base. Also, Datalog is, from a syntactic standpoint, a subset of Prolog, and a Datalog program can be parsed and executed by a Prolog interpreter.

XSB~\citep{XSB} is an implementation of Datalog; it is a complete Prolog system conforming to the ISO standard, and on top of that it further supports the integration of tabled predicates and non-tabled predicates. XSB has the following features not found in ordinary logic programming systems including Prolog.

\begin{itemize}
	\item Complete solution derivation according to the well-founded semantics is possible through SLG resolution~\citep{SLG}.
	\item Implementation of HiLog~\citep{Hilog}, a higher-order logic programming language.
	\item Various indexing-control techniques such as Unification Factoring~\citep{UF}.
	\item Source-code availability for portability and extensibility.
\end{itemize}

Many of XSB's components are based on PSB-Prolog~\citep{SWI}, but they differ in part from the Prolog system in order to process SLG resolution and HiLog expressions. For example, because Prolog's SLD resolution~\citep{SLD} is based on depth-first search, it tends to fall into infinite loops, whereas XSB's SLG resolution can correctly evaluate almost all logic programs.

\subsubsection{Translation Rules into Datalog Programs}
In the language, each declarative statement based on the syntax shown in Section~\ref{sec:design} is translated as a clause of a Datalog program, and an authorization query is evaluated as a query against this program. The terms related to Datalog are defined as follows. A literal $P$ has a predicate name and a series of arguments. An argument takes a constant term or a variable term. A clause is written as $P_0\leftarrow P_1,\ldots,P_n$ and is composed of one head literal on the left side of ``$\leftarrow$'' and a list of body literals on the right side. In this implementation, each predicate defined in the syntax is translated into a Datalog literal as follows. The left side of ``$\rightarrow$'' is the description in the language, and the right side is the translated Datalog literal. In this translation, the predicate in the $fact$ of a declarative statement or authorization query becomes the predicate of the Datalog literal, and its object is translated as the argument of the Datalog literal. The $fact$ after \textbf{if} is likewise turned into a Datalog literal by the above method and then used as the conditional clause of the Datalog clause.

\begin{align*}
&\textbf{Policy specifies}\ E\ \textbf{tagged}\ A\ \rightarrow\ \texttt{tagged(E,A)}\\
&\textbf{Policy specifies}\ A_{1}\ \textbf{inherits}\ A_{2}\ \rightarrow\ \texttt{inherits(}A_{1}\texttt{,}A_{2}\texttt{)}\\
&\textbf{Policy specifies}\ E_{1}\ \textbf{tagged}\ A_{1}\ \textbf{is permitted to}\ O\ E_{2}\ \textbf{tagged}\ A_{2}\\
&\qquad\rightarrow\ \texttt{permitted(}E_{1}\texttt{,}A_{1}\texttt{,O,}E_{2}\texttt{,}A_{2}\texttt{)}\\
&\textbf{Policy specifies}\ E_{1}\ \textbf{tagged}\ A_{1}\ \textbf{is forbidden to}\ O\ E_{2}\ \textbf{tagged}\ A_{2}\\
&\qquad\rightarrow\ \texttt{forbidden(}E_{1}\texttt{,}A_{1}\texttt{,O,}E_{2}\texttt{,}A_{2}\texttt{)}\\
&\textbf{Policy specifies}\ E_{1}\ \textbf{tagged}\ A_{1}\ \textbf{moves to}\ E_{2}\ \textbf{tagged}\ A_{2}\ \textbf{with}\ O\\
&\qquad\rightarrow\ \texttt{moves(}E_{1}\texttt{,}A_{1}\texttt{,O,}E_{2}\texttt{,}A_{2}\texttt{)}
\end{align*}

\section{Method of Expressing Access Control Models}
\label{sec:Expression}
This chapter presents methods for expressing various widely used access control models using the language. First, it explains that, rather than directly describing each access control rule as in the conventional manner, expressing them as a program using variables and subroutines makes it possible to describe the relationships among elements and frequently occurring authorization procedures concisely; it then shows concrete examples of describing access control rules using this expression method.

\subsection{Hierarchization of Elements and Structuring of Authorization Procedures}
This section presents a method for simplifying policy description by hierarchizing the elements used for access control and structuring the relationships among elements as subroutines.

Hierarchization of elements is a method for simplifying policy description that appears in various access control models; RBAC, DTE, Chinese Wall~\citep{Brewer:36295}, and others can be viewed as one way of grouping and hierarchizing access control elements. For example, RBAC is a model that groups access control rules by assigning them per role and, by further defining a hierarchical structure among roles, defines a method of collectively inheriting groups of access control rules. Similarly, DTE is a model that groups access subjects and access targets as domains and types, respectively, and, by defining access control rules between them, flexibly controls the granularity of access control.

Also, structuring the relationships among elements is a method that subroutinizes a particular authorization procedure in advance to describe it concisely; for example, TBAC~\citep{Banerjee:2005p5185} and the access control model for CSCW systems by Ahmed et al.~\citep{CSCW} abstractly structure particular authorization procedures as Authorization Steps and Activity Templates, respectively. By this, repeatedly using a subroutinized authorization procedure in transaction processing and the like reduces the amount of policy description and improves the overview.

In the language, as shown below, by fixing the access subject and its attribute as variable terms and the other terms as constant terms, and by designating the access subject using a conditional clause, the access subjects to which the relevant access control rule applies can be designated collectively.

\begin{align}
&\textbf{Policy specifies}\ x\ \textbf{tagged}\ p\ \textbf{is permitted to}\ O_{1}\ E_{1}\ \textbf{tagged}\ A_{1}\nonumber\\
&\qquad\qquad\qquad\qquad\qquad \textbf{if}\ x\ \textbf{tagged}\ A_{4}\\
&\textbf{Policy specifies}\ x\ \textbf{tagged}\ p\ \textbf{is permitted to}\ O_{2}\ E_{2}\ \textbf{tagged}\ A_{2}\nonumber\\
&\qquad\qquad\qquad\qquad\qquad \textbf{if}\ x\ \textbf{tagged}\ A_{4}\\
&\textbf{Policy specifies}\ x\ \textbf{tagged}\ p\ \textbf{is permitted to}\ O_{3}\ E_{3}\ \textbf{tagged}\ A_{3}\nonumber\\
&\qquad\qquad\qquad\qquad\qquad \textbf{if}\ x\ \textbf{tagged}\ A_{4}
\end{align}

Eq.~(1) gives, to the access subject $x$, on the condition that the attribute attached to it is $A_{4}$, a rule that permits the operation $O_{1}$ on the target $E_{1}$ tagged with attribute $A_{1}$. Similarly, Eqs.~(2) and (3) give, on the condition that the attached attribute is $A_{4}$, a rule that permits the operation $O_{2}$ on the target $E_{2}$ tagged with attribute $A_{2}$ and a rule that permits the operation $O_{3}$ on the target $E_{3}$ tagged with attribute $A_{3}$. After grouping these multiple access control rules under attribute $A_{4}$ by Eqs.~(1), (2), and (3), by attaching attribute $A_{4}$ to a particular access subject $E_{4}$ as in Eq.~(4) below, the grouped Eqs.~(1), (2), and (3) are expanded again, as Eqs.~(5), (6), and (7), to the concretized access subject.

\begin{align}
&\textbf{Policy specifies}\ E_{4}\ \textbf{tagged}\ A_{4}\\
&\textbf{Policy specifies}\ E_{4}\ \textbf{tagged}\ A_{4}\ \textbf{is permitted to}\ O_{1}\ E_{1}\ \textbf{tagged}\ A_{1}\\
&\textbf{Policy specifies}\ E_{4}\ \textbf{tagged}\ A_{4}\ \textbf{is permitted to}\ O_{2}\ E_{2}\ \textbf{tagged}\ A_{2}\\
&\textbf{Policy specifies}\ E_{4}\ \textbf{tagged}\ A_{4}\ \textbf{is permitted to}\ O_{3}\ E_{3}\ \textbf{tagged}\ A_{3}
\end{align}

At this time, Eqs.~(5), (6), and (7) are not directly described rules but rules newly generated by assigning and expanding the grouped rules; by attaching attribute $A_{4}$ to a concrete access subject as in Eq.~(4), it is possible to repeatedly generate new rules.

In addition, by defining transition conditions among the grouped attributes using the predicate ``\textbf{moves to}'' as below, the language abstractly structures a particular authorization procedure.

\begin{align}
&\textbf{Policy specifies}\ x\ \textbf{tagged}\ p\ \textbf{moves to}\ x\ \textbf{tagged}\ A_{5}\ \textbf{with}\ O_{3}\nonumber\\
&\qquad\qquad\qquad\qquad\qquad \textbf{if}\ x\ \textbf{tagged}\ A_{4}\\
&\textbf{Policy specifies}\ x\ \textbf{tagged}\ A_{5}\ \textbf{is permitted to}\ O_{4}\ E_{5}\ \textbf{tagged}\ A_{6}
\end{align}

Eq.~(8) shows a rule by which, when attribute $A_{4}$ is attached to access subject $x$, its attribute transitions to $A_{5}$ triggered by $O_{3}$. By this transition, the attribute of access subject $x$ changes to attribute $A_{5}$. Eq.~(9) defines a rule that permits $O_{4}$ to an access subject tagged with attribute $A_{5}$ for the access target $E_{5}$ tagged with attribute $A_{6}$. By these, the access rules grouped by $A_{4}$ further transition by Eq.~(8) to Eq.~(9), thereby abstractly structuring the authorization procedure leading from Eqs.~(1), (2), and (3) to Eq.~(9). Also, by concatenating multiple transition rules, even a complex authorization procedure such as transaction processing can be expressed concisely.

\subsection{Examples of Describing Access Control Rules}
This section shows examples of actually expressing various widely used access control models using the description method shown in the previous section. Here we take up RBAC as an example of hierarchization and TBAC as an example of subroutinizing an authorization procedure.

\paragraph{RBAC.}
RBAC is a model that describes a policy by associating multiple access control rules with an attribute abstracted as a role; because it is naturally easy to apply to organizational forms such as enterprises, it is widely and generally used. In the example below, Eq.~(10) gives an access subject tagged with the role \textsc{Manager} the privilege to \textsc{Read} an access target tagged with \textsc{Billing\_Information}, and Eq.~(11) gives an access subject tagged with the role \textsc{Floor\_Leader} the privilege to \textsc{Read} an access target tagged with \textsc{Accounting\_Information}. Also, by Eq.~(12), \textsc{Manager} inherits the privileges given to \textsc{Floor\_Leader}. By these rules, even for the same access subject, the privileges given can be changed by the role attached.

\begin{align}
&\textbf{Policy specifies}\ x\ \textbf{tagged}\ p\ \textbf{is permitted to}\ \textsc{Read}\ y\ \textbf{tagged}\ q\nonumber\\
&\qquad\qquad\qquad\quad \textbf{if}\ x\ \textbf{tagged}\ \textsc{Manager},\nonumber\\
&\qquad\qquad\qquad\qquad\ y\ \textbf{tagged}\ \textsc{Billing\_Information}\\
&\textbf{Policy specifies}\ x\ \textbf{tagged}\ p\ \textbf{is permitted to}\ \textsc{Read}\ y\ \textbf{tagged}\ q\nonumber\\
&\qquad\qquad\qquad\quad \textbf{if}\ x\ \textbf{tagged}\ \textsc{Floor\_Leader},\nonumber\\
&\qquad\qquad\qquad\qquad\ y\ \textbf{tagged}\ \textsc{Accounting\_Information}\\
&\textbf{Policy specifies}\ \textsc{Manager}\ \textbf{inherits}\ \textsc{Floor\_Leader}
\end{align}

\paragraph{TBAC.}
TBAC is a model that, after grouping authorization procedures as Authorization Steps (AS), determines an authorization decision by designating, in addition to the access subject, access target, and operation content, the AS name and which step within the AS one is at. In this model, various relationships can be designated for each element during transaction processing. In the example below, Eq.~(13) gives an access subject in the \textsc{Prepare\_Phase} the privilege to \textsc{Request\_to\_Prepare} \textsc{DB1}, which is \textsc{Not\_Prepared}, and Eq.~(14) gives an access subject in the \textsc{Commit\_Phase} the privilege to \textsc{Commit} \textsc{DB1}, which is \textsc{Prepared}. Also, Eq.~(15) designates that an access subject in the \textsc{Prepare\_Phase} transitions to the \textsc{Commit\_Phase} triggered by \textsc{DB1\_Prepared}. By these rules, the privileges given can be changed according to the step of transaction processing.

\begin{align}
&\textbf{Policy specifies}\ x\ \textbf{tagged}\ p\ \textbf{is permitted to}\nonumber\\
&\qquad\qquad \textsc{Request\_to\_Prepare}\ \textsc{DB1}\ \textbf{tagged}\ q\nonumber\\
&\qquad\qquad\qquad \textbf{if}\ x\ \textbf{tagged}\ \textsc{Prepare\_Phase},\nonumber\\
&\qquad\qquad\qquad\quad \textsc{DB1}\ \textbf{tagged}\ \textsc{Not\_Prepared}\\
&\textbf{Policy specifies}\ x\ \textbf{tagged}\ p\ \textbf{is permitted to}\ \textsc{Commit}\ \textsc{DB1}\ \textbf{tagged}\ q\nonumber\\
&\qquad\qquad\qquad \textbf{if}\ x\ \textbf{tagged}\ \textsc{Commit\_Phase},\nonumber\\
&\qquad\qquad\qquad\quad \textsc{DB1}\ \textbf{tagged}\ \textsc{Prepared}\\
&\textbf{Policy specifies}\ x\ \textbf{tagged}\ \textsc{Prepare\_Phase}\ \textbf{moves to}\nonumber\\
&\qquad\qquad x\ \textbf{tagged}\ \textsc{Commit\_Phase}\ \textbf{with}\ \textsc{DB1\_Prepared}
\end{align}

\section{Evaluation of the Policy Description Language}
\label{sec:Evaluation}
This chapter presents the results of the experiments conducted to evaluate the usefulness of the language and their discussion. First, to confirm the validity of the language, we actually composed, using the language, the policy used in SELinux, and by confirming the identity of the response contents at the time of authorization decisions, we demonstrated that the language operates logically correctly. After that, to evaluate the expressiveness of the language, we quantitatively compared, from multiple viewpoints, the amount of description required to compose a policy of the same content between SELinux's policy description method and the language, and confirmed that the language's expression method can express an actual policy concisely. Finally, based on these results, we discuss the advantages and disadvantages of the language.

\subsection{Configuration of the Experimental System}
The experimental system was implemented on a computer with a Pentium~4 (3.0~GHz) and 1~GB of memory, and its OS environment uses Debian/GNU Linux with SELinux-related packages added. The main software versions are libselinux1 (2.0.65-5), selinux-basics (0.3.5), and selinux-policy-default (2:0.0.20080702-6). The experimental system obtains the addresses of the sidtab and avtab that exist in the kernel space, and uses pointers of their hash arrays to obtain the entire contents of the tables. At the same time, it expresses, in the language, the information of all the elements lined up in the obtained avtab, and prepares an authorization decision mechanism loaded into Datalog, ready for queries by combinations of source context, target context, and security class. The measurement program was written in the C language, and the results were recorded as a text file by converting the context and class information into name strings.

\subsection{Validity Evaluation Experiment by Comparison of Response Contents}
\label{sec:Validity}
This section presents the results of evaluating the identity of the response contents between the SELinux policy and the same policy composed in the language. In the experiment, using all the elements of the sidtab, we called $context\_struct\_compute\_av()$ in a round-robin manner over the security classes to obtain the AV (Access Vector), and then queried Datalog with the same combination of source context, target context, and security class, and evaluated whether the obtained AV matched the one returned by $context\_struct\_compute\_av()$.

In this experiment, a total of 15,100,162 queries generated using the combinations of all elements of the sidtab and all security classes were evaluated; of these, 875,644 were those for which both responded with the same AV, and 14,216,639 were those whose response contents matched in the sense that neither returned any AV. As a result, the response contents were identical between SELinux and the language for a total of more than 99\%, namely 15,092,283 queries. Also, for the remaining approximately 0.05\%, namely 7,879 queries, the response contents did not match.

\begin{table}[tb]
	\caption{Comparison of source-code line count and listing page count}
	\centering
	\begin{tabular}{lcc}
		\toprule
		Evaluation target & Source-code lines & Listing pages\\
		\midrule
		SELinux policy source           & 6,524 & 133\\
		Proposed language (C-List style) & 335   & 95\\
		Proposed language (ACL style)    & 356   & 96\\
		\bottomrule
	\end{tabular}
	\label{table:function}
\end{table}

\subsection{Expressiveness Evaluation Experiment by Comparison of Description Amount}
\label{sec:Description}
This section presents the results of comparing the amount of description required when expressing the SELinux policy in the language, by the number of source-code lines required to compose the policy and the number of pages when listed (Table~\ref{table:function}).

In the language, using the expression method shown in Section~\ref{sec:Expression}, even access control rules of the same content allow various expressions, so in this experiment we described them in two particular ways. The first is, like a capability list~\citep{Capability} (C-List), a description method in which the source context as the attribute of the access subject is designated as a constant term, while the target context and security class as attributes of the access target and the AV as the operation content are designated as variable terms, and the contents of those variable terms are concretely specified within the conditional clause. The second is, like an access control list~\citep{Saltzer} (ACL), a description method in which the target context as the attribute of the access target is designated as a constant term, while the source context and security class as attributes of the access subject and the AV as the operation content are designated as variable terms, and the contents of those variable terms are concretely specified within the conditional clause.

In this experiment, the policy described in 6,524 source-code lines by SELinux's description method was described in 335 lines by the C-List-style description method using the language, and in 356 lines by the ACL-style description method. Also, as for the number of pages when listed, the SELinux policy source file had 133 pages, whereas when described in the language it became 95 pages and 96 pages for the C-List style and ACL style, respectively.

\subsection{Discussion of the Experimental Results}
In the validity evaluation experiment, the response contents matched for a total of more than 99\% between the queries decided as access-permitted and those decided as access-denied, but some cases were confirmed in which the response contents did not match. The cause of this is that queries were made against contexts dynamically generated during system operation, and this depends on how the arguments are passed at the time of the authorization decision. Therefore, this resolution needs to be devised outside the authorization decision mechanism; on the other hand, when one presupposes that SELinux's original authorization decisions are valid with respect to the policy description, it can be said that the policy description by the language and the authorization decisions based on it were themselves also valid. Accordingly, this experimental result showed that the language, based on logic programming, can describe an actual policy logically correctly.

As an application of this, by applying the features of logic programming to access control, one can consider, for example, flexibly verifying a policy by queries using variables and describing that verification result as a condition of access control; controlling, by meta-rules, how new rules are added or existing rules are changed; and having the authorization decision mechanism change the arguments used for the authorization decision according to the situation.

In the expressiveness evaluation experiment, we showed that the description method using the language can compose the SELinux policy with a small amount of description; this is because many of the designations needed to define an access control rule are collected into the conditional clause, with the result that the contents of the conditional clause for each rule become large. Therefore, SELinux's description method, which directly designates the necessary elements, makes it easier to intuitively grasp the content when each rule is viewed individually. On the other hand, in the language's description method, because access control rules are described in a somewhat collective manner through hierarchization and subroutinization, the readability when each rule is viewed individually is not so good, but the overview as the whole policy is improved.

Therefore, an expression that directly designates elements, like SELinux's policy description method, and an expression that uses variable terms and conditional clauses, like the language, have different characteristics, so their advantages and disadvantages are considered to depend on the phase of use. For example, when one wants to strictly describe a small number of access control rules that have no relevance to other rules, the former description method is suitable; and when many access procedures shared by various authorization decisions exist and one wants to designate them collectively, or when the total number of access control rules becomes huge and flexible structuring is needed, the latter is more suitable.

\section{Conclusion}
\label{sec:Conclusion}
In this paper, we proposed a policy description language that, by expressing access control rules as a logic program, enables the inheritance of attributes and descriptions structured as subroutines, and we presented a method for resolving the complexity related to policy description that becomes an issue when applying multi-layer defense effectively. The language is designed in its essentials by defining the syntax, the formal semantics, and the inference rules of the declarative statements for describing each access control rule and of the authorization queries for querying the policy composed as a set of declarative statements; it is implemented using Datalog, a subset of Prolog. Also, in this paper, after presenting a method for structurally describing concrete access control models using the language, we constructed an experimental system that actually described the SELinux policy, and evaluated the validity and the expressiveness of authorization decisions. In the validity evaluation experiment, we demonstrated that, when SELinux's authorization decisions are presupposed to be valid, the authorization decisions by the language show logically correct responses and are valid. Furthermore, in the expressiveness evaluation experiment, we showed that the description method using the language can compose the SELinux policy with a small amount of description, and from both experimental results we discussed the usefulness of the language.

In the future, we will first examine methods to actually realize the applications of the language shown as expected effects in Section~\ref{sec:purpose}. We will also examine methods for configuring an authorization decision mechanism that performs access control of an actual system using the language and for making it cooperate with the access control mechanisms of an OS and the like. In doing so, in order to configure the authorization decision mechanism itself securely, and also from the viewpoint of improving performance, we plan to embed the Datalog processing system into the kernel.

\subsubsection*{Acknowledgments}
We are grateful to the editorial committee members and the anonymous reviewers for their valuable advice.

\bibliographystyle{unsrtnat}
\bibliography{refs}

\end{document}